\documentclass[]{spie}  %>>> use for US letter paper
\usepackage{amsmath,amsfonts,amssymb}
\usepackage{graphicx}
\usepackage{float}
\usepackage[colorlinks=true, allcolors=blue]{hyperref}
\usepackage{multirow}

\title{Multichannel-GaAsP-photomultiplier-based fiber bundle ISM-STED microscope}

\author[a]{Marcus Babin}
\author[a]{Simone Fabian}
\author[a,b]{Michael Reibe}
\author[c]{Lukas Spantzel}
\author[c]{Michael Börsch}
\author[a]{Erik Beckert}
\affil[a]{Institute for Applied Optics and Precision Engineering - Fraunhofer IOF, Albert-Einstein-Straße 7, Jena, Germany}
\affil[b]{Friedrich Schiller University Jena, Institute of Applied Physics, Albert-Einstein-Straße 15, Jena, Germany}
\affil[c]{Single-Molecule Microscopy Group, Jena University Hospital, Nonnenplan 4, Jena, Germany}

\authorinfo{Further author information: (Send correspondence to Marcus Babin)\\E-mail: marcus.babin@iof.fraunhofer.de}

\begin{document} 
\maketitle

\begin{abstract}
The usage of a GaAsP  (gallium arsenide phosphide) photomultiplier for microscopical imaging allows the evaluation of low-light luminescent objects. We designed a setup for collecting a confocal microscopic image signal, which is divided into 14 equal-sized input channels. The division is achieved with a beamsplitter and two fiber bundles consisting of seven fibers each. Re-imaging the confocal pinhole by such a densely packed fiber bundle permits the utilization of a photon re-localization approach to overcome the optical resolution limit. The center fiber creates a real-time image, while the outer fibers enable a higher-resolution image via an image scanning microscope (ISM) signal calculation. 
%Additionally, stimulated emission depletion (STED) suppresses the fluorescence below and above the image plane.
The fiber bundles are enclosed in a fused silica capillary and are drawn out to create one solid fiber bundle. During the drawing process, the fiber bundles are tapered down to an outer diameter size of 400 µm, with each fiber having a less than 0.3 Airy unit diameter. For the photomultiplier interface, all fibers of both fiber bundles are integrated into a v-groove array, with each fiber representing a detection input, which is followed by projection optics for imaging onto the multichannel detector. The resulting confocal super-resolution microscope is suitable for the application of time-correlated single photon counting (TCSPC) techniques such as fluorescence lifetime imaging (FLIM), time-resolved anisotropy, or Förster resonance energy transfer (FRET) imaging.
\end{abstract}


\section{INTRODUCTION}
\label{sec:intro}

Single-photon detectors enable the recording of faintly luminescent light sources by increasing the signal-to-noise ratio of a limited number of incoming photons. This sensitivity allows the detection of a single fluorescent molecule and, therefore, to reach the ultimate chemical sensing limit. Using a pinhole-based single detector in a conventional confocal laser-scanning microscope \cite{GREGOR201974}, imaging fluorophore-tagged single biomolecules like a protein, a lipid molecule or a nucleic acid in the complex environment of a cell is possible. The fluorescence properties in regular microscopic images mainly comprise brightness or photon counts per pixel, respectively. Advanced confocal microscopes provide additional information from a fluorophore, i.e., FLIM, spectral shifts of excitation and emission wavelengths, fluorescence anisotropy, diffusion constant, and distance-dependent fluorophore-fluorophore interaction at the nanometer scale like FRET. However, the accurate localization of a single molecule in the microscopic image is restricted by the fundamental optical resolution limit to about half the wavelength of the detected photons. 

Different microscopy approaches have been developed to overcome the optical resolution limit for measuring the position of a fluorescent molecule. Single-photon counting cameras enable super-localization microscopies like photoactivated localization microscopy (PALM, STORM, PAINT, etc.) to achieve a position accuracy in the range of ten nanometers. Confocal stimulated emission depletion (STED) microscopy can reach a similar resolution for specific fluorescent molecules of nanoparticles. Structured illumination microscopy (SIM, camera-based) or image scanning microscopy (ISM, based on re-imaging the confocal pinhole) provide a moderate increase in position resolution about a factor of two compared to widefield fluorescence microscopy. Still, image recording is much faster, and other fluorescence properties besides the brightness information can be measured simultaneously. 

\begin{figure} [ht]
   \begin{center}
   \begin{tabular}{c} %% tabular useful for creating an array of images
   \hspace{-0.2cm}
   \includegraphics[height=4.05cm]{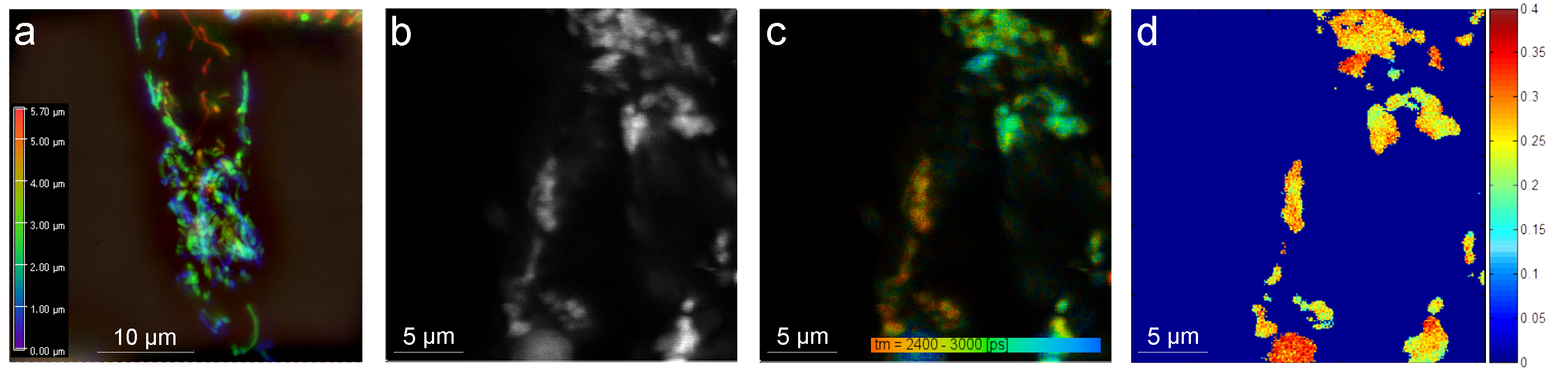}
   \end{tabular}
   \end{center}
   \caption[]
   { \label{fig:STED_example} 
%STED microscopy images of the mitochondria of a HEK293 cell, recorded with varying imaging techniques: \textbf{a}) superresolution image (SIM), \textbf{b}) FLIM and steady-state anisotropy images
Microscopic images of the fluorescently tagged FoF1-ATP synthases in the mitochondria of a living HEK293T cell, recorded with varying imaging techniques:  \textbf{a}) superresolution image (3D-SIM, different colors represent the height position of the mitochondria with respect to the cover glass bottom),  \textbf{b}) confocal fluorescence intensity image,  \textbf{c}) false-colored confocal FLIM image with a lifetime color bar, and  \textbf{d}) false-colored steady-state confocal anisotropy image with fluorescence anisotropy color bar. 
\cite{foertsch2017imaging}.}
\end{figure}

The implementation of an array of GaAsP PMTs allows an extension of an advanced confocal microscope with the functionality of an ISM \cite{Castello2019ARA}. The ability of the GaAsP PMT to temporally resolve single photons opens up the opportunity to utilize time-dependent imaging techniques such as FLIM or FRET on the ISM. Our goal is the real-time recording of structural dynamics and molecular rearrangements of individual membrane proteins in the living cell. Specifically, we aim to image the enzyme FoF1-ATP synthase in the inner membrane of the mitochondria of human kidney cells (HEK293T). We have tagged the enzyme with a fluorophore mNeonGreen and have used the camera-based SIM approach to localize the enzyme within the inner membrane of the mitochondria in living HEK293T cells (see Fig.~\ref{fig:STED_example} a). To obtain information about the local interactions of these enzymes we have recorded FLIM and fluorescence anisotropy images in a separate set of experiments using an advanced confocal microscope (see Fig.~\ref{fig:STED_example} b). However, understanding the restructuring processes of the inner mitochondrial membranes as related to changes in the molecular environment of the membrane proteins requires simultaneous super-localization and interaction mapping of the individual FoF1-ATP synthases at work.
 
We present a fiber-based PMT-coupled optical microscope that uses two fiber bundles with seven fibers each in a rotational symmetric arrangement. Each fiber’s inputs are filtered transmitting either horizontal or vertical linear polarization coming from a system input defining pinhole. Ultimately, it is the goal of this fiber bundle-based microscope to improve the spatial resolution using ISM and the temporal resolution using PMTs for such low-light emitting fluorescently-tagged membrane proteins of mitochondria in living cells. 

\section{DESIGN AND EMPLOYED TECHNOLOGIES}
\subsection{PMT Array Integration}
\label{PMT_Array}
\begin{figure} [ht]
   \begin{center}
   \begin{tabular}{c} %% tabular useful for creating an array of images 
   \includegraphics[height=6.9cm]{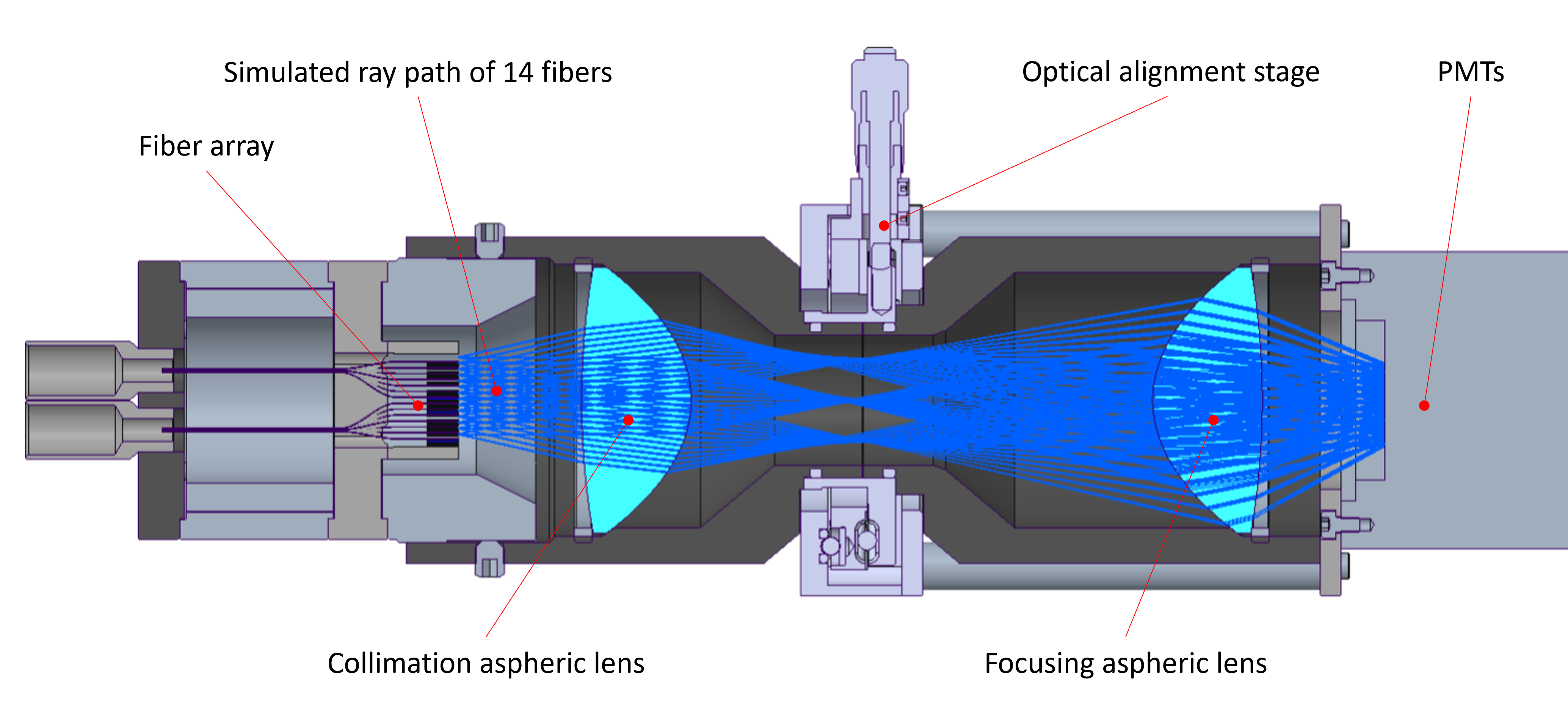}
   \end{tabular}
   \end{center}
   \caption[]
   { \label{fig:xsec_fiber_array} 
Digital rendering of the optical and the derived mechanical design, with the traced paths of the rays leaving the linear fiber array and being imaged over two aspheric lenses onto the linear array of PMTs. All optical parts are enclosed by mechanical housing elements, that assure a stable position with respect to each other.}
\end{figure}

For the employment of the fiber bundles and the PMTs, an optical setup on the basis of a 4F optical system using Fourier optics was designed to image the fiber outputs onto the PMTs input surfaces. The core elements of this design are two aspheric condenser lenses with a two-inch diameter, a numerical aperture of 0.6, and a broadband antireflection coating ranging from 350~nm to 700~nm. From this optical foundation, a mechanical design was derived to incorporate the input-giving fibers of the fiber bundle and the comparatively bulky PMTs. To match the linear array of the PMTs with the fibers of the bundles, the single fibers are spread out and similarly arranged into a linear array as the PMTs. The mechanical housing elements are coaxially positioned with location clearance fits, fine threads and precision machined end stops. Alignment of the optical parts with respect to the PMTs is ensured with a two-axis differential screw-driven alignment stage. The fiber array can be precisely aligned with machined position marks and an active alignment routine using the PMTs output when inputting a constant multimode light signal into the fiber bundles. A digital image of the mechanically integrated optical design is shown in Fig.~\ref{fig:xsec_fiber_array}.

\subsection{Confocal Pinhole Imaging}
\begin{figure} [H]
   \begin{center}
   \begin{tabular}{c} %% tabular useful for creating an array of images 
   \includegraphics[height=6.5cm]{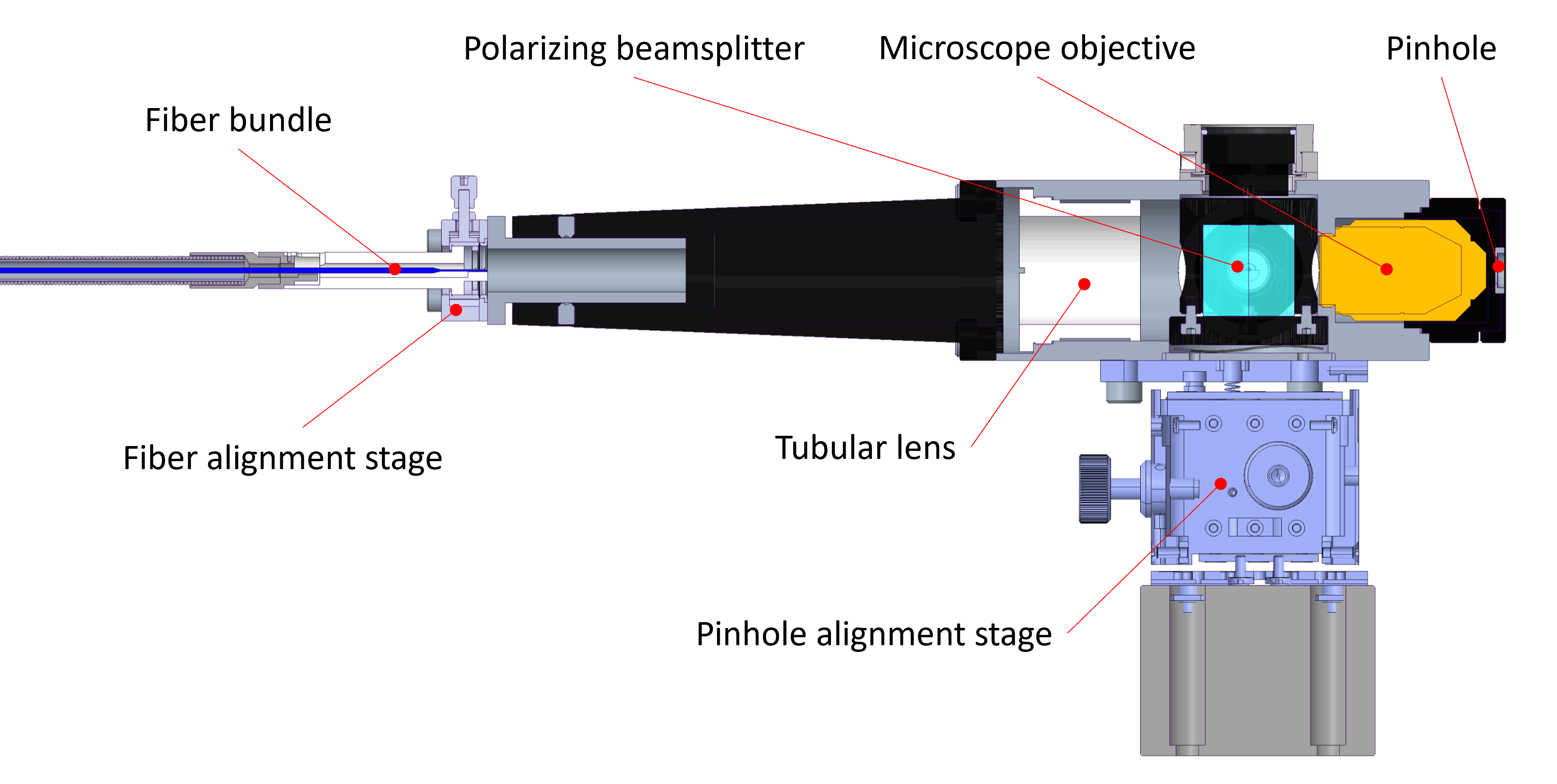}
   \end{tabular}
   \end{center}
   \caption[]
   {\label{fig:xsec_microscope} 
Digital rendering of the mechanical design of one arm of the fiber-based microscope of the beamsplitter transmitted polarization state integrating the beginning of the fiber bundle and the system input defining entry pinhole.}
\end{figure}

To engage with the signal of a conventional confocal microscope with STED microscopy, the output of the microscope is imaged onto a system entering pinhole. The signal is then magnified with a standard microscope objective and tubular lens setup. Finally, the signal is imaged onto the entry surface of both fiber bundle facets (see Fig. \ref{fig:xsec_microscope}). The image containing light is separated by its perpendicular electric field components transverse to its direction of propagation for the input of each fiber via a polarizing beamsplitter. 

The design allows the free exchange of microscope objectives with varying degrees of magnification levels within the same standard. The complete system can be aligned with an auxiliary three-axis alignment stage so that the pinhole is in focus concerning the output of a confocal microscope.

\subsection{Tapering of the Fiber Bundle}

The confocal image is scanned with two fiber bundles each consisting of seven standard silica core multimode fibers. To assemble the fiber bundle, the seven fibers were threaded inside a silica capillary and fixated. The final bundle diameter was defined by the to-be-imaged confocal microscope pinhole and was set to 400~µm so that each fiber images a 0.3 Airy unit diameter. In principle, the outer diameter can be defined freely in a fiber taper process and can be chosen according to the desired optical entry facet size. 
%During this process, the application of focused $\text{CO}_2$ laser light generates heat that leads to the glass-liquid transition of the silica material. 
A CO2 laser beam shaped as a laser ring radially heats the capillary with the internal fibers above the softening point of fused silica. 
Because of silica's strong absorption coefficient of the 10.6~µm wavelength of the  $\text{CO}_2$ laser, most of the induced energy will be transformed into internal energy \cite{nold18}. Applying an axial load will draw out the capillary and its containing fibers, causing it to be axially elongated and radially contracted, due to the Poisson effect. Before threading of the fibers, the capillary was drawn in a first taper process to the corresponding inner capillary diameter to accommodate the seven standard fiber clad diameters. 

\begin{figure} [ht]
   \begin{center}
   \begin{tabular}{c} %% tabular useful for creating an array of images 
     \includegraphics[height=6cm, angle=-90]{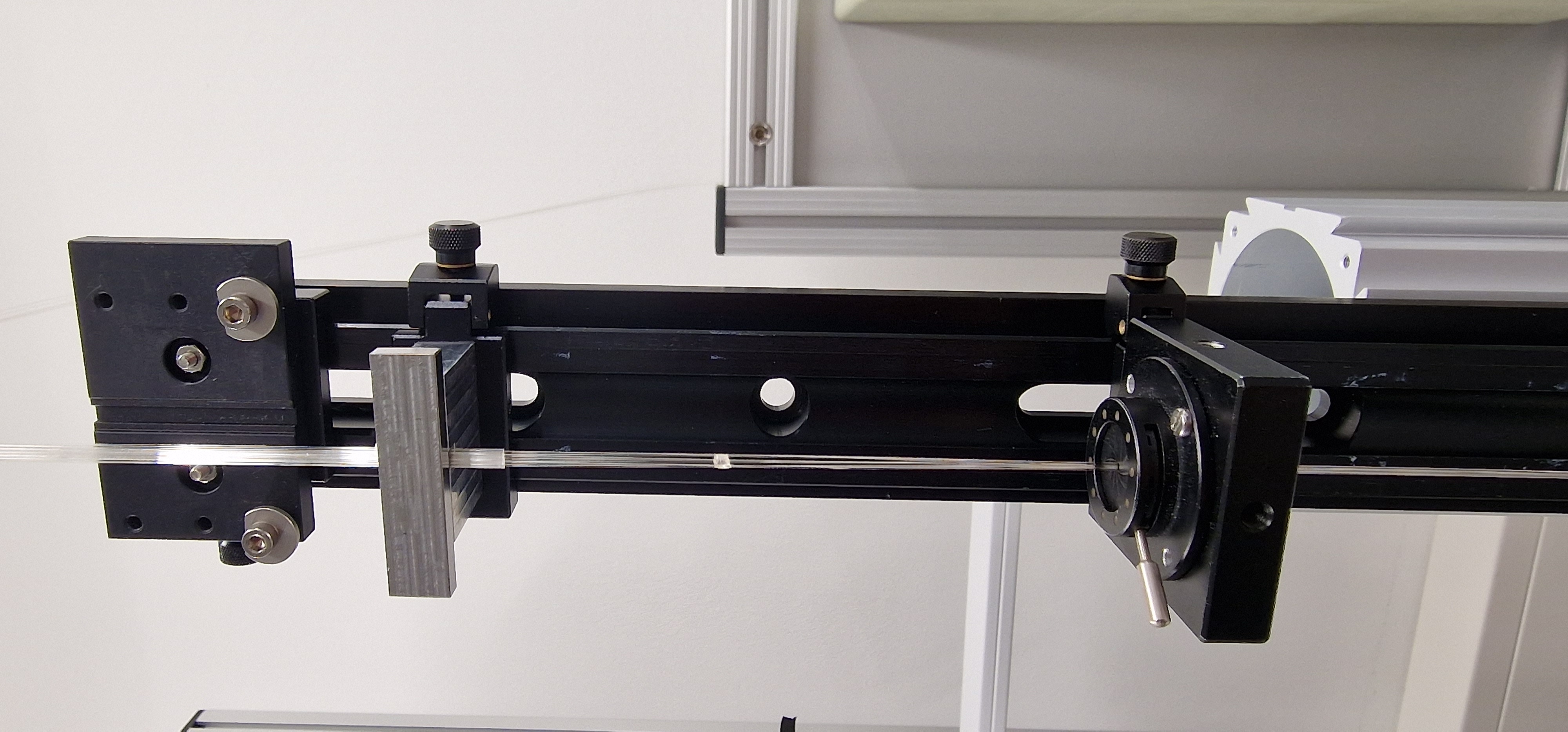}\hspace{1cm}
     \includegraphics[height=6cm, angle=-90]{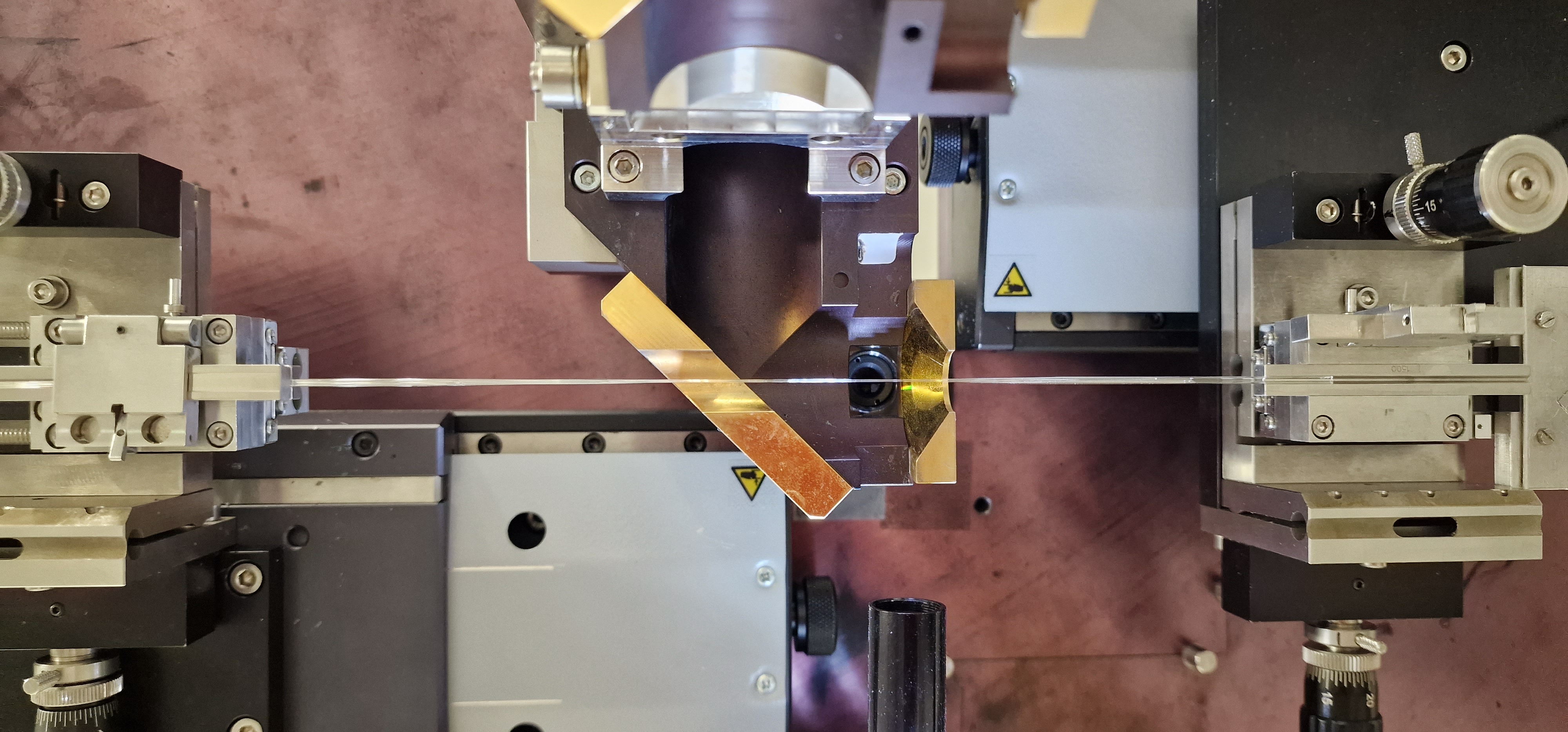}
   \end{tabular}
   \end{center}
   \caption[] 
   { \label{fig:taper} 
Preform setup to assemble seven single fibers into a silica capillary (left). The single fibers are clamped or glued into a fixed bundle to realize a strain relief. Overview of the fiber taper machine with its movements stages top and bottom gripping the fiber capillary compound with the axial redirecting and fiber focusing conical laser mirrors (right) \cite{Boehme_2009}.}
\end{figure}  

If assembled in the hexagonal close-packing of circles on a two-dimensional plane, the initial assembly inner capillary diameter would be equal to the sum of the clad diameter of three fibers, resulting in a 660~µm diameter. After threading, the fibers are fixed in place relative to the capillary (see Fig. \ref{fig:taper} (left)) and tapered in a second process as a common assembly to achieve the targeted fiber bundle diameter of 400~µm (see Fig. \ref{fig:taper} (right)).

\section{SETUP ASSEMBLY AND CHARACTERIZATION}
\subsection{Fiber Bundle Implementation}

\begin{figure} [ht]
   \begin{center}
   \begin{tabular}{c} %% tabular useful for creating an array of images 
      \includegraphics[height=4.14cm]{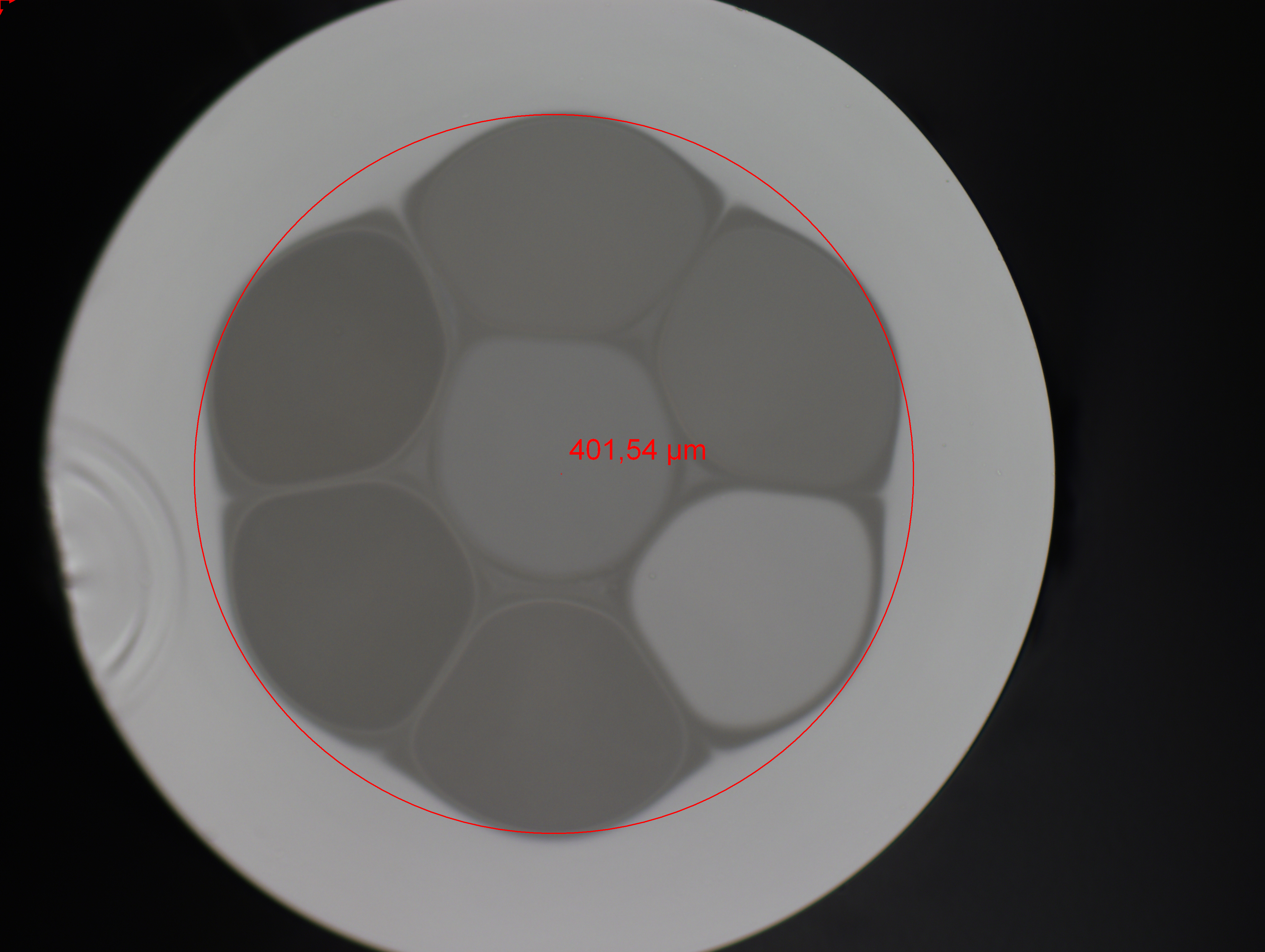}
      \includegraphics[height=4.14cm]{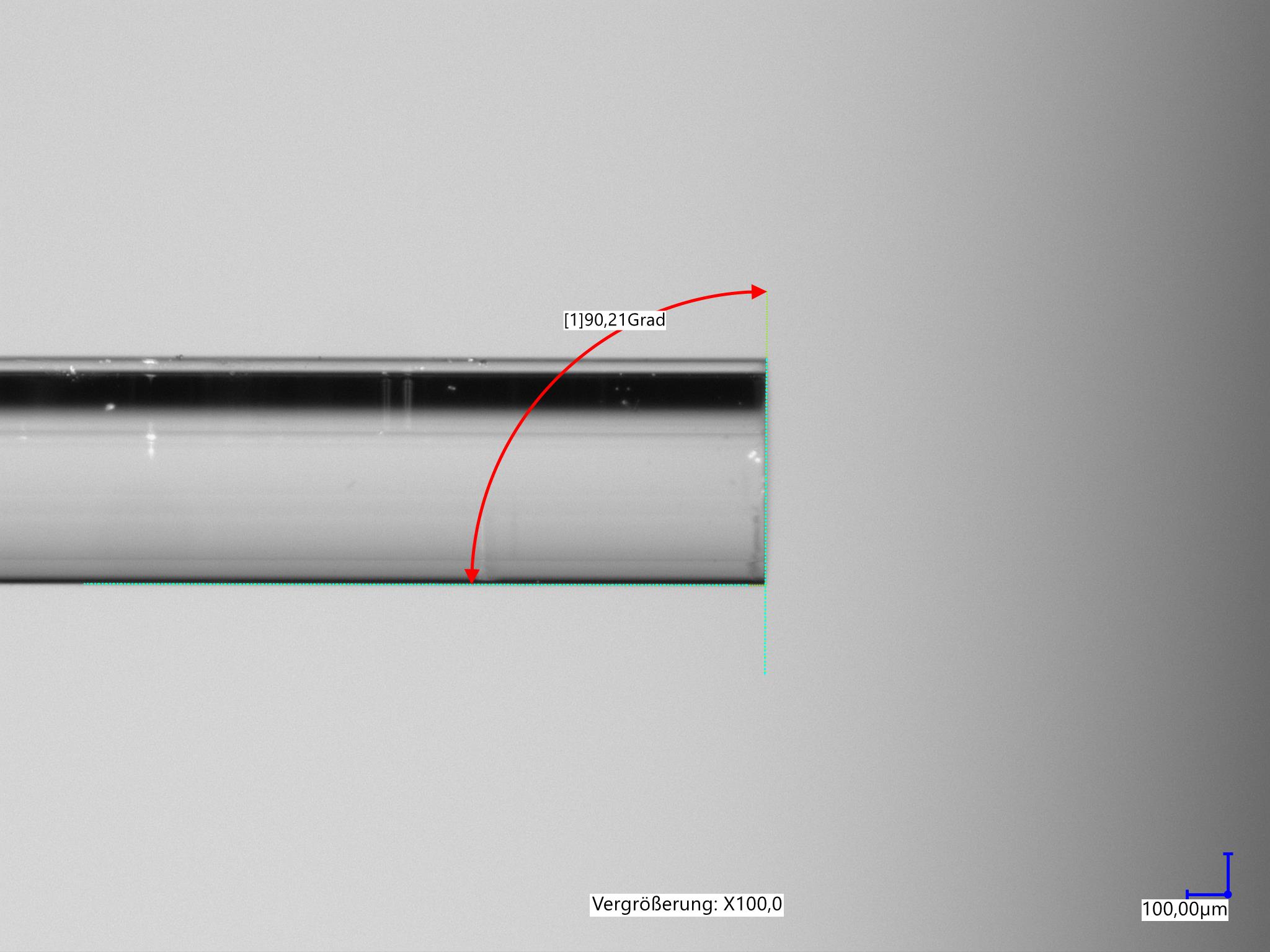}
      \includegraphics[height=4.14cm]{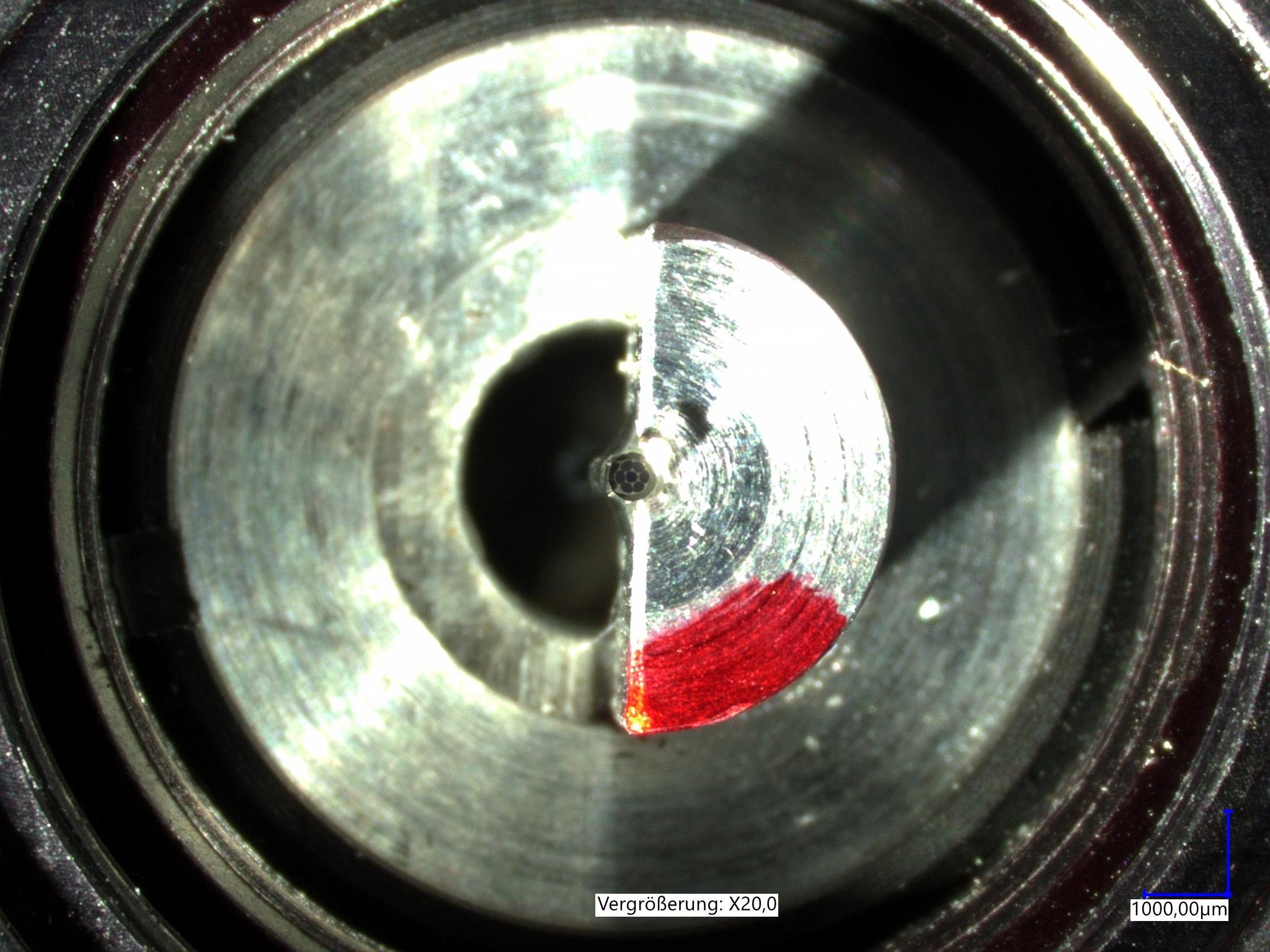}
   \end{tabular}
   \end{center}
   \caption[]
   { \label{fig:fiber_bundle} 
Microscope images of the fiber bundle and its mechanical integration. Image of the front facet of the fiber bundle with a \(\sim\) 401.5~µm bundle diameter (left), side view of the bundle end with a facet cleave angle of \(\sim\) 90.2$^{\circ}$ (center), and an image of the bundle integrated into the fiber alignment stage (right).}
\end{figure}

%During the taper process, the drawing speed controls the gradient of the taper and ultimately the resulting diameter. The laser power and focus control the total amount of heat that will be introduced and the impacted volume respectively. 
A suitably selected laser power and the set speed difference of both motor axes, which hold the capillary filled with fibers, lead to a melted fiber bundle with a defined reduction of both the outer diameter of the capillary and the inner fibers. 
Micrometer control of the final tapered diameter requires precise control of the temperature and drawing speed. To achieve a coaxial integration of the fiber bundle with respect to the optical axis, the fiber bundle will be placed in an appropriately sized v-groove geometry according to the outer diameter of the drawn-out fiber enclosing capillary. This integration approach requires a constant outer diameter of the capillary for the length of the v-groove. Consequently, both the fiber bundle diameter of 400~µm and the constant outer capillary diameter are target parameters of the tapering process, where at first, during a down-taper step, the fiber bundle diameter will converge to the targeted 400~µm and then in a second drawing step, an elongation of over 15~mm is achieved. During the down-taper, the drawing speed will be ramped up and will be kept constant once the targeted diameter is reached. Critical during this procedure is that the temperature at the heating zone, the area where the volumetric forming takes place, is kept constant. After tapering and drawing, the fiber bundle will be cleaved and its resulting facet polished to optical quality. The results of one produced fiber bundle are shown in Fig.~\ref{fig:fiber_bundle}, with its end facet shown in Fig.~\ref{fig:fiber_bundle} (left), a side view of the bundle shown in Fig.~\ref{fig:fiber_bundle} (center) and the v-groove integrated bundle in Fig.~\ref{fig:fiber_bundle} (right). 

\subsection{Fiber Array Implementation}

\begin{figure} [H]
   \begin{center}
   \begin{tabular}{c} %% tabular useful for creating an array of images 
      \includegraphics[height=5.3cm]{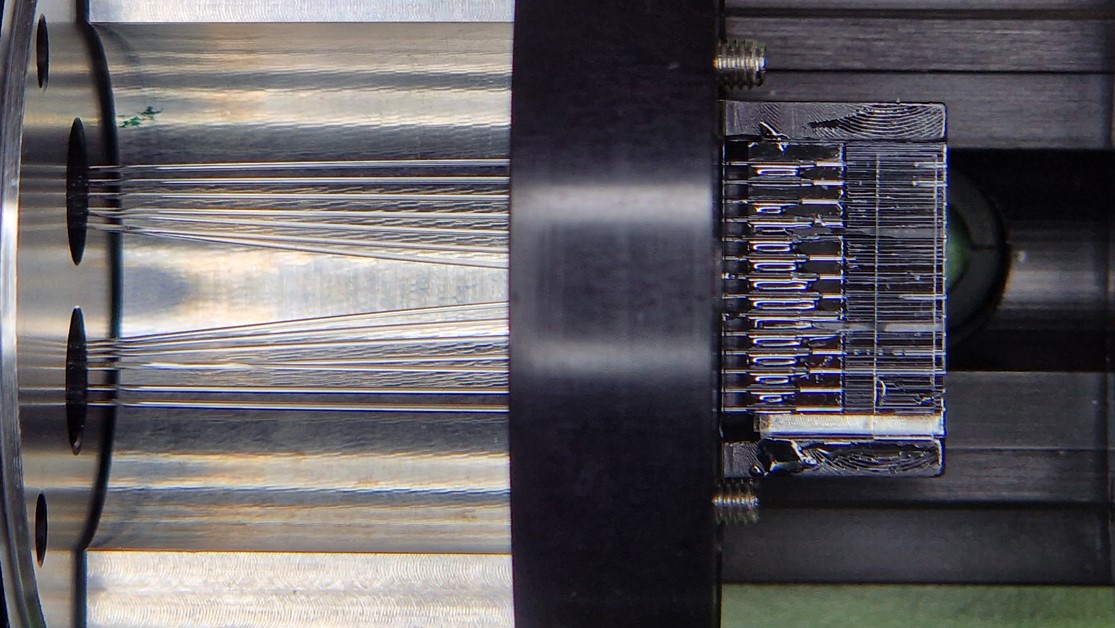}
     \includegraphics[height=5.3cm]{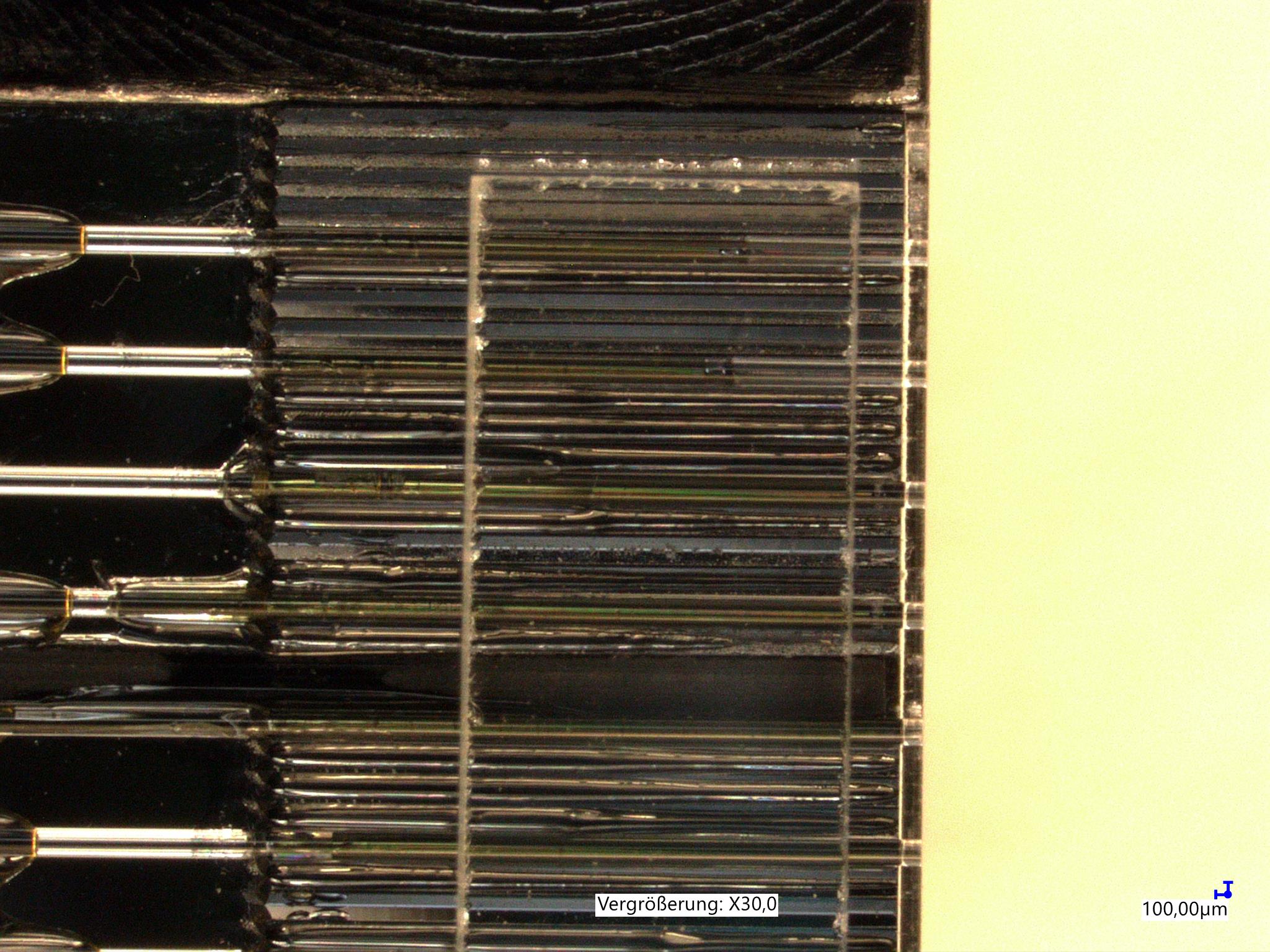}
   \end{tabular}
   \end{center}
   \caption[]
   { \label{fig:fiber_array} 
Images of the fiber integration into silicon v-grooves. Image of the fiber array assembly integrated into the mechanical housing as shown in figure \ref{fig:xsec_fiber_array} (left). Microscope image of the fibers glued into v-grooves showing the micrometer exact positioning of the fiber end facets (right).}
\end{figure}

The single fibers that are leading out of the opposing end of the tapered fiber bundle are integrated into lithographically etched silicon v-grooves \cite{harold} via microscopic alignment and low volume adhesive application (see Fig. \ref{fig:fiber_array} (right)). It is critical for the in Sec. \ref{PMT_Array} presented imaging system, that the fibers are parallel with respect to each other, have an equal pitch, and end at the same distance to the first lens. The silicon v-grooves assure parallelism and pitch accuracy, while a precision machined temporarily fixed end stop guarantees a closely matching fiber end distance. A cover glass is used to exert pressure onto the fibers and as a result, become self-centered inside their respective v-groove. In the final step, the v-groove integrated fibers from the two fiber bundles are glued in via the silicon array inside the v-groove carrier mount. The full fiber array assembly planarly integrated on the array mount is depicted in Fig. \ref{fig:fiber_array} (left).

% \begin{figure} [H]
%    \begin{center}
%    \begin{minipage}{6.75cm}
%        \includegraphics[height=5cm]{figures/Fasern8-14-KCh4gr3_Bild1ÜS.jpg}
       
%        \vspace{0.25cm}
%        \includegraphics[height=5cm]{figures/Fasern1-7-KCh4gr4_Bild1ÜS.jpg} 
%      \end{minipage} \hspace{0.05cm}
%    \begin{minipage}{6.75cm}
%        \includegraphics[width=10.25cm, angle=90]{figures/Fiber_Array_Top.jpg}
%    \end{minipage}
%    \end{center}
%    \caption[]
%    { \label{fig:fiber_array} 
% Images of the fiber integration into silicon v-grooves. Microscope images of the fibers glued into v-grooves showing exact positioning of the fiber end facets (left). Fiber array assembly integrated into the mechanical housing as shown in figure \ref{fig:xsec_fiber_array} (right).}
% \end{figure} 

\section{CONCLUSION AND OUTLOOK}

\begin{figure} [H]
   \begin{center}
   \begin{tabular}{c} %% tabular useful for creating an array of images 
        \includegraphics[height=5.3cm]{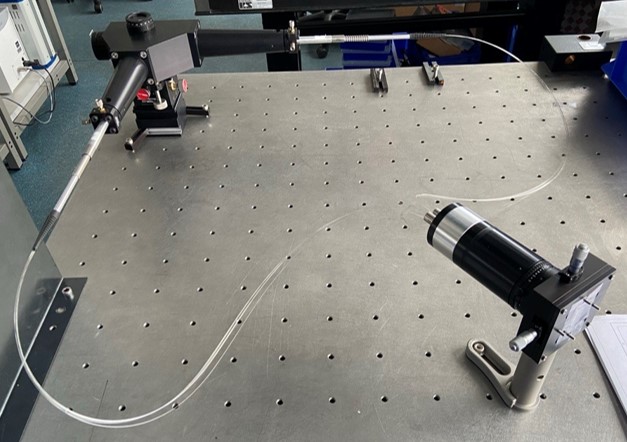}\hspace{0.1cm}
        \includegraphics[height=5.3cm]{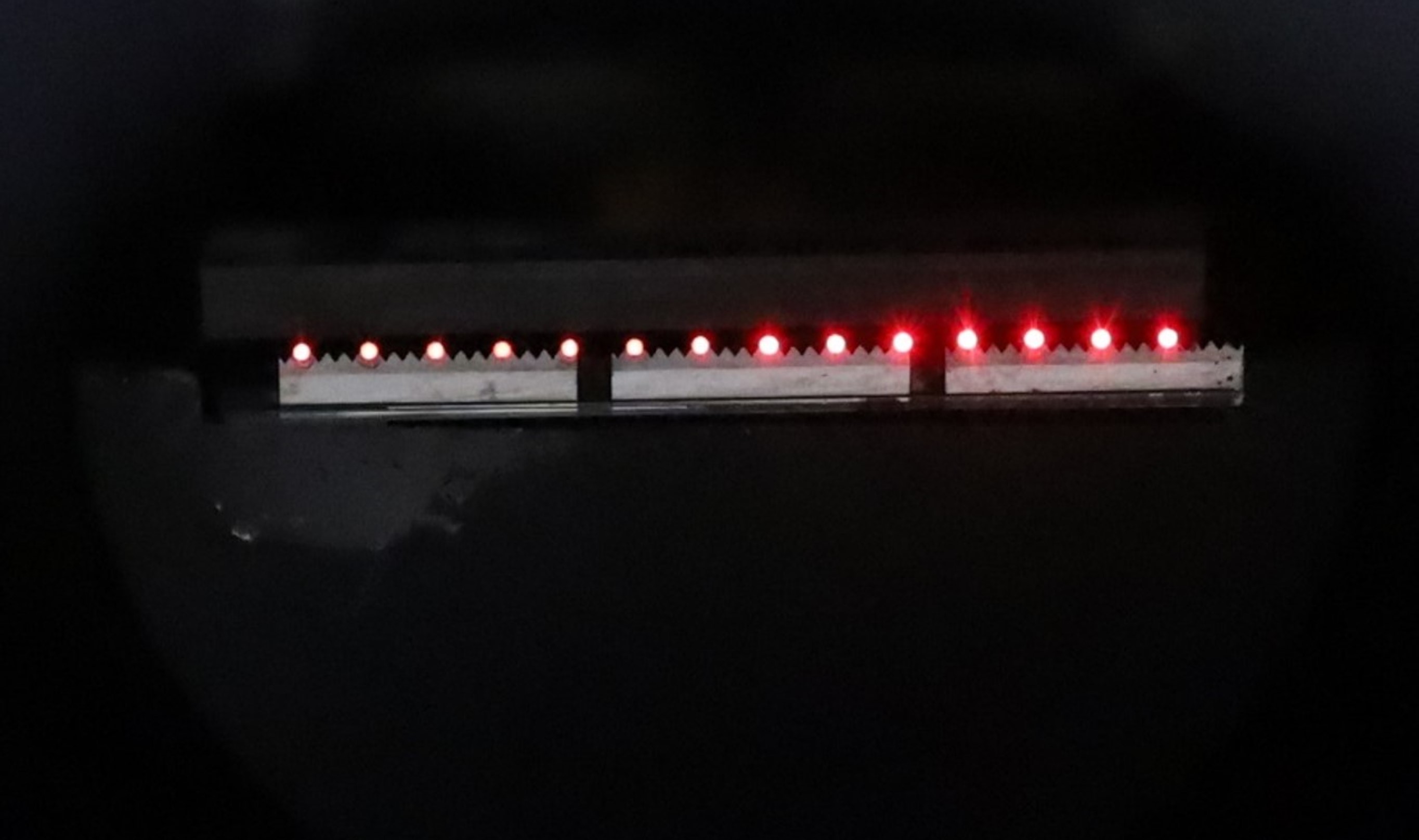}
   \end{tabular}
   \end{center}
   \caption[]
   { \label{fig:setup} 
Assembled microscope setup with fiber bundle assembly unit and the fiber array-PMT unit on an optical breadboard (left). Recorded exit signals from the single fibers assembled in the v-groove array (right).} 
\end{figure}

The two sub-assemblies of the fiber bundle and array were ultimately fully assembled with their optical elements and alignment stages on top of an optical breadboard (see Fig.~\ref{fig:setup} (left)). For initial characterization, a 680~nm laser was used to fully illuminate the input pinhole. After calibration of both fiber bundles in- and output to their respective optical components, the exit signals were recorded (see Fig.~\ref{fig:setup} (right)). In this characterization setup, the signals reveal the arbitrarily chosen orientation of the linearly polarized input laser beam with respect to the polarizing beamsplitter in front of the fiber bundles. 

Next, we will assign the 14 fiber outputs to individual detector channels of an 16-channel GaAsP PMT (PML-16-GaAsP, Becker Hickl, Berlin, Germany) connected to the corresponding TCSPC electronics (PCIe card SPC 180NX, Becker Hickl, Berlin, Germany) for FLIM recording. Fiber-dependent optical power losses will be characterized, the timing performance of each channel for FLIM will be determined, and the overall imaging quality will be assessed and compared using a custom-designed confocal piezo-scanning microscope dedicated for time- and polarization-resolved single-molecule detection.

In its final implementation step, the fiber bundle ISM microscope will be integrated into an existing advanced confocal STED-FLIM-FRET microscope (Expert line, Abberior instruments, Göttingen, Germany) at the Jena University Hospital. On a separate exit port of this microscope, we will combine ISM for an improved lateral resolution with the z-STED mode for an improved axial resolution. This will result in an isotropic 3D resolution increase for the microscopic images. With minimized photon losses for FLIM and anisotropy or FRET measurements, the structural dynamics of the essential membrane enzymes of the cellular powerhouses, i.e., the mitochondria in living cells, will come into focus.

\acknowledgments      
 
The authors thank Steffen Böhme and Dr. Steffen Trautmann of the Fraunhofer Institute for Applied Optics and Precision Engineering, Jena, Germany for their work in the development and implementation of the taper machine and its input software, and the contribution of the design of the optical image system. The authors gratefully achnowledege funding by the State of Thuringia (ACP Explore project within the ProExcellence initiative ACP2020, to Mi.B. and E.B.) and by the Deutsche Forschungsgemeinschaft (grant INST 1757/25-1 FUGG, project number 411346541, to Mi.B.).

% References
\bibliography{report} % bibliography data in report.bib
\bibliographystyle{spiebib} % makes bibtex use spiebib.bst

\end{document}